 *************************************************************************
 *  All figures for the preprints from Nuclear Physics Group of          *
 *  Drexel University are available through the ftp.                     *
 *  The procedure is : ftp at                                            *
 *        einstein.physics.drexel.edu (129.25.1.120)                     *
 *        anonymous                          (as username)               *
 *        <Return>                           (as password)               *
 *        cd pan_files                       (go to the subdirectory)    *
 *************************************************************************

 For the paper entitled
      "Chaos" in Nuclear High Spin Spectroscopy
 The filenames for the three figures are goefig1.ps and goefig2.ps
 and goefig3.ps

\documentstyle[preprint,aps]{revtex}
\begin{document}
\def\bra{\langle}
\def\ket{\rangle}
\draft
\preprint{\today}
\title{ ``Chaos" in Nuclear High Spin Spectroscopy}
\author{Yang Sun$^{(1)}$, Da Hsuan Feng$^{(1)}$, Hsi-Tseng Chen$^{(2)}$ and Hua
Wu$^{(3)}$}
\address
{$^{(1)}$Department of Physics and Atmospheric Science, Drexel University \\
Philadelphia, Pennsylvania 19104 \\
$^{(2)}$Department of Physics, Chung-Yuan Christian University \\
Chung-Li, Taiwan, ROC \\
$^{(3)}$Department of Physics and Astronomy, McMaster University \\
Hamilton, Ontario, Canada L8S 4M1}

\maketitle

\begin{abstract}
The Projected Shell Model with zero-, two- and four-quasiparticle
configurations
is used to investigate the level statistics, i.e. chaoticity, of high spin
spectroscopy. The model can describe many high spin phenomena and with the
present configuration space has sufficient number of levels for statistical
analysis. It is found that the degree of chaoticity has a sensitive dependence
on the classification of levels in question, and that it steadily increases
with excitation energy and angular momentum.
\end{abstract}

\newpage
Level-spacing statistics, {\it a la} Random Matrix Theory \cite{Me.67} is an
important tool to study nuclear spectroscopy. In the past decade, it has also
gradually emerged as one of the workable definitions of order and chaos for a
quantum system, such as nucleus \cite{CHAOS}. Although there is no commonly
accepted definition of ``quantum chaos" \cite{NOCHAOS}, there is no doubt that
level-spacing statistics does produce two limiting distributions: Poisson and
GOE (Gaussian Orthogonal Ensemble), which, for a large class of models,
correspond to ordered and chaotic classical motions, respectively.
It would be interesting to know ``in which windows of the parameter
space (energy, angular momentum, parity {\it etc.}) will the distribution of
the nuclear levels be GOE-like". If this question can be answered, and
if one could firmly establish that GOE distribution is indeed a manifestation
of ``quantum chaos", then one can further investigate questions such as
``how does a nuclear system become chaotic". This Letter will address the
former question.

About a decade ago, Hag, Pandey and Bohigas \cite{IPB.82}
began to search for such windows in the nuclear
system. However, they were immediately faced with a difficulty which persisted
to this date: The number of available data is well below what is required for a
statistical analysis \cite{IPB.82,EXP}. To remedy this,
they performed the statistical analysis on groups of
states collected from different nuclei, the nuclear data ensemble.
However, Abul-Magd and Weidenm\"uller realized early on that the outcome
of the statistical analysis can be dependent on the classification and
selection of the data. They showed that one can obtain different distributions
for different groups of the same set of data \cite{AW.85}. Still, this
important idea is somewhat compromised since they, faced with the
aformentioned shortage of data, were unable to separate them according to
parity
and angular momentum.

With the empirical statistical analysis stalled, to answer the above mentioned
question will, within the forseeable future, rely on theoretical models. Of
course, this line of research will be intimately linked to the question of how
well such models can reproduce the known data and can construct sufficient
number of levels. Recently, Alhassid and Vretenar
\cite{AV.92} and Wu, Feng and Valli\`eres \cite{WFV.90} made significant steps
toward this direction by using, respectively, the Interacting Boson Model
(IBM) and the Fermion Dynamical Symmetry Model (FDSM) \cite{WFG.94}.
While the latter concentrated only on the question of dynamical
symmetry breaking in low energy spectroscopy, the former also studied the
statistical behavior of the spectroscopy at high spins. It deserved mentioning
that none of these approaches has any realistic nuclei in mind and that the
suitability of the IBM for high spin states is still very much a topic of
current investigation.

It is therefore highly desirable to have a model which can reproduce well the
spectroscopy of high spin states to carry out the level-spacing statistics.
Straightforward implementation of the spherical shell model is naturally out of
question for the heavy systems. The ambitious approach (MONSTER)
developed by the Tuebingen group \cite{SG.87} could in principle be used, but
since for practical reasons the configurations are restricted to 2-qp states,
it may not be sufficient to study the statistical behavior of levels in higher
excitation energies. The FDSM \cite{WFG.94} can also in principle allow such
studies be carried out, and its implementation is currently underway. Quite
recently, a model called the Projected Shell Model (PSM), proposed in the late
1970s \cite{HI.80} by the Munich group, has undergone extensive development
\cite{HS.91,SE.94}. A code for this model was developed and successfully used
to study a range of high spin phenomena in the rare earths. It is thus a
practical shell model approach to describe the deformed heavy systems.

Since the PSM has been discussed in several publications, and is the subject
matter of a forthcoming review article \cite{HS.94}, we shall only touch upon
the relevant features here. Roughly speaking, unlike the conventional shell
model, the PSM begins with the deformed (Nilsson-type \cite{An.78}) single
particle basis. Its advantage over the conventional shell model is that the
important nuclear correlations are easily taken into account and the
configuration space is manageable, thus making the shell model treatment for
heavy systems possible. Also, it deserves to be emphasized that the results
obtained from the PSM can be interpreted in simple physical terms.
Such a shell model basis violates the rotational symmetry, but it can be
restored by the standard angular momentum projection technique. Pairing
correlation is included by a successive BCS calculation for the Nilsson states.
Thus, the shell model truncation is carried out within the quasiparticle states
with the vacuum $|\phi>$.

The angular momentum projected wave function for the PSM is given by
$| I M > ~=~ \sum_{\kappa} f_{\kappa} \hat P^I_{MK_\kappa} | \varphi_{\kappa}
>$, where $\kappa$ labels the basis states. Here we shall assume axial symmetry
in the intrinsic states $| \varphi_{\kappa} >$. Thus $K$ is a good quantum
number. Using the Tamm-Dancoff-Approximation \cite{RS.80}, the basis states $|
\varphi_{\kappa} >$ for a doubly even system are spanned by
\begin{eqnarray}
\left\{\;\; |\phi >, \;\;\; \alpha^\dagger_{i} \alpha^\dagger_{j} |\phi >,
\;\;\; \alpha^\dagger_{i} \alpha^\dagger_{j} \alpha^\dagger_{k}
\alpha^\dagger_{l} |\phi >, \;\;\; ...~ \right\} ,
\label{baset}
\end{eqnarray}
where $\{\alpha, \alpha^\dagger \}$ are the quasiparticle annihilation and
creation operators for the vacuum $|\phi >$. To restore the rotational
symmetry,
these intrinsic state $| \varphi_{\kappa} >$ will be acted on by the projection
operator $\hat P^I_{MK}$ \cite{RS.80}.  This will generate states of good
angular momentum. For example, $\hat P^I_{MK}|\phi>$ will describe the
ground-state (g-) band.  Thus, the model has the inherent flexibility that by
including higher order multi-quasiparticle states in eq. (\ref{baset}), one can
reach levels of higher excitation energy and angular momentum. In fact, if one
were to construct all possible configurations in eq. (\ref{baset}), then one
will reach the exact shell model space. However,
for reproducing the yrast properties,
a basis with 60 low-lying configurations is sufficient
\cite{HS.91}.

In this work, the single particle space is sufficiently large and consists of
three major shells: i.e. N = 4, 5 and 6 (N = 3, 4 and 5) for neutrons
(protons). The multi-quasiparticle basis
states of eq. (\ref{baset}) are constructed by both the normal and
the abnormal parity orbitals. We should also point out that the 4-qp states are
built from a neutron and a proton pairs.  Such states will in general lie
lower in energy than the like-particle (protons and neutrons) 4-qp states.
The latter states are not included in the present work. Also,
the size of the basis states of eq. (\ref{baset})
is determined by using the (unprojected quasiparticle) energy cut-off of 5 MeV
for both 2- and 4-qp states. Within this energy window, there will be about 200
2-qp states from the N = 5 and 6 neutron and N = 4 and 5 proton major shells.
Likewise, about 500 4-qp states based on those 2-qp states are constructed.
So the dimension of the basis is about 700.
It was pointed out by \AA berg \cite{AA.88} that the 6-qp states will begin to
play a role at the (super-deformed high rotating)
particle-hole excitation energy in the
vicinity of 4 MeV. The possible effect of the missing 6-qp and the
like-particle 4-qp configurations will be discussed later.  Also, the following
hamiltonian is used
\cite{HI.80}:
\begin{equation}
\hat H = \hat H_0 - {1 \over 2} \chi \sum_\mu \hat Q^\dagger_\mu
\hat Q^{}_\mu - G_M \hat P^\dagger \hat P - G_Q \sum_\mu \hat
P^\dagger_\mu\hat P^{}_\mu .
\label{hamham}
\end{equation}
This is essentially the well-known Pairing plus Quadrupole Hamiltonian
\cite{BK.68} which is known to describe not only the nuclear ground-state
properties but the region at finite temperature as well \cite{ER.93}.
In addition, there is also a quadrupole pairing term whose
importance was recently once again demonstrated \cite{SWF.94}. Yet, despite the
simplisity, this hamiltonian has worked surprisingly well in predicting
various high spin phenomena \cite{HS.91,SE.94,SWF.94,SFW.94}. The interaction
strengths are determined as follows: the quadrupole interaction strength $\chi$
is adjusted so that the known quadrupole deformation $\epsilon _2$ is obtained
from the Hartree-Fock-Bogoliubov self-consistent procedure \cite{Lamm.69}. The
monopole pairing strength $G_M$ is adjusted to the known energy gap
$G_M =  \left[ 20.12\mp 13.13 \frac{N-Z}{A}\right] \cdot A^{-1}$, where the
minus (plus) sign is for neutrons (protons). The quadrupole pairing strength
$G_Q$ is assumed to be proportional to $G_M$ and the proportional constant is
fixed at 0.16 in this work. Diagonalization of the Hamiltonian of eq.
(\ref{hamham}) in the truncated projected basis of eq. (\ref{baset}) is
equivalent to mixing states with different $K$ quantum numbers. For a given
angular momentum $I$ and parity $\pi$, one will finally obtain a set of
energy levels $\{E_i\}$. The highest state obtained after diagonalization
for each $I$ can reach an excitation region of 7 MeV above the yrast line.
The total number of obtained levels is typically 460 (650) for $I = 4\hbar$
($I = 10\hbar$).

To illustrate the physics, we have chosen a typical backbender $^{164}$Er. This
is a well deformed nucleus with a purely rotational ground band. It was
previously shown \cite{HS.91} that with the same model,
one can well describe the high
spin phenomena in the yrast region of this and many other deformed nuclei.
Hence there is no loss of generality to choose this nucleus as an example. We
shall first focus our discussion on the level statistics as a function of
excitation energy for a given angular momentum ($I = 10\hbar$)
and positive parity. In
the top part of Fig.1, the statistics for the entire set of levels for this
spin and parity are given. From both the nearest
level-spacing distribution $P(X)$ and
the $\Delta_3(L)$ statistics we see that the distribution lies
somewhere between the Poisson and GOE limits. As we divide them into groups
according to excitation energies, clear pictures of Poisson and GOE
distributions emerge. In the middle part of Fig.1, the group of levels are
those
up to 2 MeV excitation energy above the yrast line. Here both $P(X)$ and
$\Delta_3(L)$ are manifestly Poisson-like. When only states of higher
excitation energies are included (4 $\rightarrow$ 6 MeV), GOE statistics
appears (see bottom part of Fig.1).

It should be emphasized that the mixing of $K-$states by the two-body residual
interactions play a key role in understanding the above results. The
projected quasiparticle basis forms a spin/parity group, with each state
having a definite $K$. Hence diagonalization will mix the $K-$states and the
degree of the $K-$mixing can differ for the different excitation and spin
regions. Clearly, there will be strong mixing between bands lying close in
energy and with the same symmetry, which is the case for the region of high
excitation energies. On the other hand, at low excitations, $K$ is
approximately a good quantum number because bands are well separated. Hence
our analyses suggest that the level-spacing statistics could be an indication
of the amount of $K-$mixing for a certain group of levels in well deformed
nuclei, namely, the Poisson
statistics will signal a weaker and the GOE a stronger mixing of the
$K-$states. This can also explain why the top part of Fig.1 is an intermediate
situation since the group of states in question includes both stronger and
weaker $K-$mixing ones.

To enhance the above idea, we shall show in Fig.2 four data groups. These are
groups of the calculated $I^{\pi} = 10^+$ levels of $^{164}$Er, each with a
different energy limit at the top (i.e. 0 $\rightarrow$ 3,
0 $\rightarrow$ 4, 0 $\rightarrow$ 5 and 0 $\rightarrow$ 6 MeV). Together
with the middle part of Fig.1, we see that there is a smooth transition from
Poisson to GOE, thus allowing us to speculate that with increase in energy, the
strong mixing component will eventually dominate the statistics, resulting in a
pure GOE picture.

We now turn our attention to the dependence of the statistics on nuclear
rotation, or explicitly, angular momentum. It is well-known from the
phenomenological argument that the Coriolis force tends to mix the symmetry of
a rotating system. With larger angular momentum, the force will of course
increase as well. In the microscopic model, it can be shown that at
higher spins, level density is high already near the yrast region
\cite{HS.91}. Therefore, it is easy to conjecture that level-spacing
distribution will steadily approach the GOE limit for sufficiently rapid
rotation.
Our calculations here do support this conjecture (see the plots
of the left row in Fig.3).

In a recent publication \cite{AV.92}, Alhassid and Vretenar used the IBM with
one broken pair to demonstrate that the GOE distribution is best manifested
in the vicinity of the crossing between the ground and the lowest 2-qp bands.
Beyond it, the Poisson statistics appears to be restored. This is in direct
contradiction to the results we obtained here. To resolve this discrepency, we
extracted from eq. (\ref{baset}) only the vacuum and the 2-qp states built by
the $i_{13\over 2}$ neutrons. This is thus a caricature of the model used in
ref. \cite{AV.92}. The size of this basis is 70. It should be mentioned that
with such a basis, the yrast band of $^{164}$Er can be adequately described and
the backbending at spin 16$\hbar$ is reproduced exactly \cite{HS.91}. The
results for this basis are shown on the right row of plots in Fig.3. As
expected, the results are consistent with ref. \cite{AV.92}, i.e. the
distribution is GOE-like at the band crossing region and approaches
Poisson for higher
spins. These results suggest that the conclusion of Alhassid and Vretenar
could be due, at least in part, to the lack of 4-qp components in their
calculation. It also implies that our results (the left row of Fig.3)
could be more GOE-like when the like-particle 4- as well as 6-qp
configurations are included.

In the present study, the angular momentum is treated fully quantum
mechanically. However, the particle number is conserved only on the average.
Therefore, our conclusions contain the assumption that statistics
of the neighboring nuclei, especially in the well deformed region, should
not differ much in character. Although there is no evidence against this
assumption, it should nevertheless be checked in later studies.

To conclude, we have extended the application of the Projected Shell Model,
which has been successfully used to describe the high spin
spectroscopy of the yrast region,
to the higher excitation regions. The level-spacing statistics was carried
out by using up to 700 0-, 2- and 4-qp levels, reaching as high as 7 MeV above
the yrast line. We showed that the resulting statistics has a strong
dependence on how the levels are classified. Steady transition
from Poisson-like to GOE-like was found
as one moves from low
to high excitation energy and/or from low to high spin. It
appears that we have thus answered the question, stated in the beginning of
this Letter, of ``in which windows of the parameter space the distribution of
the nuclear levels will become GOE-like". Clearly, we see that the presence
of the GOE spectroscopy seems to satisfy much of our intuitive understanding of
chaos. However, for a deeper understanding,
one must answer the next question of ``how a nuclear system becomes chaotic".
Yet, the final link between GOE and chaos, a central question of the
mysterious field of ``quantum chaos" remains to be made.

\acknowledgments
Useful discussions with P. Ring, T. von Egidy and S. \AA berg are acknowledged.
Yang Sun is most grateful to the College of Arts and Science of Drexel
University for the provision of a research fellowship. This work is partially
supported by the United States National Science Foundation.

\baselineskip = 14pt
\bibliographystyle{unsrt}

\begin{figure}
\caption{ Statistics for the unfolded energy levels of $I^{\pi} = 10^+$. The
plots on the left row show the nearest-neighbor level spacing distribution
$P(X)$ and those on the right the $\Delta_3$ statistics. For both rows,
the dotted (dashed) curves correspond to Poisson (GOE) distribution. 1)
Top: the entire set of levels (650). 2) Middle: Levels from
the yrast state up to 2 MeV. 3) Bottom: Levels from 4 to 6 MeV.}
\label{figure.1}
\end{figure}
\begin{figure}
\caption{
Statistics for the unfolded energy levels of $I^{\pi} = 10^+$ for different
level classifications. The descriptions of the curves are the same as in Fig.1.
}
\label{figure.2}
\end{figure}
\begin{figure}
\caption{
The $\Delta_3$ statistics as a function of angular momentum.  The dotted
(dashed) curves correspond to Poisson (GOE) distribution. 1) Left row:
statistics for the entire set of levels obtained from the whole basis
at each spin. 2) Right row:
statistics for the restricted basis. For details, see the text.}
\label{figure.3}
\end{figure}

\end{document}